\documentclass[prl,twocolumn,showpacs,preprintnumbers]{revtex4}
\usepackage{amsfonts}
\usepackage{amsmath}
\usepackage{amssymb}
\usepackage{graphicx}
\usepackage{dcolumn}
\usepackage{bm}
\usepackage{mathcomp}

\begin{document}

\begin{abstract}
We investigate the coherence properties of an array of one-dimensional Bose gases with short-scale phase fluctuations. The momentum distribution is measured using Bragg spectroscopy and an effective coherence length of the whole ensemble is defined. In addition, we propose and demonstrate that time-of-flight absorption imaging can be used as a simple probe to directly measure the coherence-length of 1D gases in the regime where phase-fluctuations are strong. This method is suitable for future studies such as investigating the effect of disorder on the phase coherence.

\end{abstract}
\title[1D gases]{Momentum-resolved study of an array of 1D strongly phase-fluctuating Bose gases}
\author{N.~Fabbri*$^{,1}$, D.~Cl\'ement$^{1,\dag}$, L.~Fallani$^{1}$, C.~Fort$^{1,2}$ and M.~Inguscio$^{1}$}
\affiliation{$^{1}$LENS and Dipartimento di Fisica e Astronomia, Universit\`a di Firenze, and INO-CNR,\\ via N. Carrara 1, I-50019 Sesto Fiorentino (FI), Italy.\\$^{2}$C.N.I.S.M. UdR di Firenze, Via Sansone 1, I-50019 Sesto Fiorentino (FI), Italy.}
\maketitle

Physics of one-dimensional (1D) systems attracts a great interest both on the theoretical and the experimental side. Recent progress in nanotechnology allowed to implement 1D systems in a variety of fields, from inorganic and organic superconductors \cite{inorganicsupercondgiamarchichem}, carbon nanotubes and nanowires \cite{auslaender2005} to spin chains and ladders \cite{spins}, as well as cold atomic systems \cite{laburthetolra2004,kinoshita2004paredes2004}. All these systems belong to the universality class of interacting quantum fluids known as Luttinger liquids \cite{haldane1979}, whose properties strongly differ from their 3D counterparts. For instance, quantum and thermal fluctuations are strongly enhanced by the reduced dimensionality, their knowledge giving access to key quantities characterizing the system \cite{giamarchibook}. Their presence can drastically alter the properties of the systems, such as in superconductive disordered nanowires where they can lead to the formation of phase-slip centers \cite{michotte2003vodolazo2003}.

In the context of cold atoms, both phase and density fluctuations of 1D systems have been studied in the last years \cite{esteve2006,hofferberth20072008,richard2003,dettmer2001}. In particular, phase coherence has been investigated by monitoring interference between two different 1D gases \cite{hofferberth20072008} and by observing density modulations \cite{dettmer2001} or the response to light scattering \cite{richard2003} in elongated 3D quasi-condensates. Nevertheless, in all these realizations transverse trapping frequencies hardly exceed a few kHz and are typically of the order of the chemical potential and the temperature. Reaching the regime of strongly interacting 1D systems would further enhance the presence of quantum and thermal fluctuations. This can be achieved when atoms are loaded in 2D optical lattices allowing for much stronger transverse confinements \cite{laburthetolra2004,kinoshita2004paredes2004}. Yet in the latter case one obtains a large number of 1D tubes for which techniques like \cite{hofferberth20072008,dettmer2001} cannot be implemented to study the phase coherence properties since tube-averaging washes out the response signal.

In this work we investigate the axial coherence properties of an array of strongly phase-fluctuating 1D Bose gases and we suggest time-of-flight (TOF) imaging as a probe of the coherence length. In our case, thermally-induced phase-fluctuations dominate and drastically reduce the coherence length ($2 L_\phi$) of the system compared to a 3D Bose-Einstein condensate (BEC) \cite{cazalilla2004petrov2000}. We use Bragg spectroscopy with large momentum-transfer \cite{stenger1999} to measure the momentum distribution and directly evaluate $L_\phi$ \cite{richard2003}. In addition, we verify that direct mapping of the momentum distribution into coordinates-space via absorbtion imaging after TOF is an effective probe of the phase-fluctuations. We demonstrate that in our range of parameters these two techniques yield the same results.

\begin{figure}[h!]
\includegraphics[width=1.0\columnwidth]{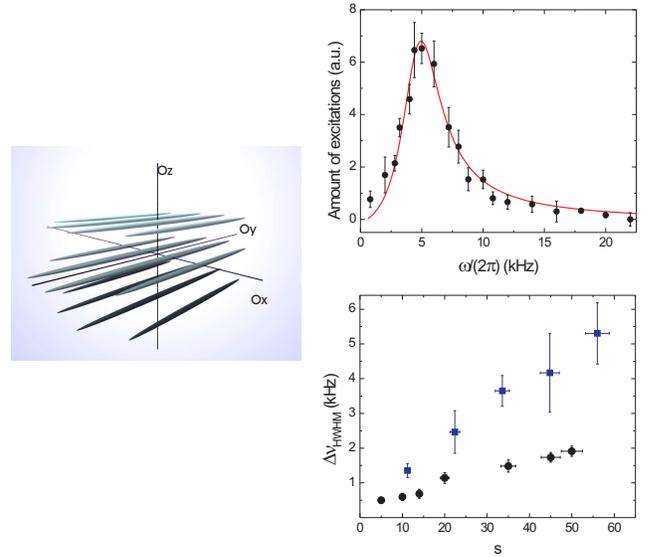}
\caption{(a) Schematic view of the two-dimensional array of one-dimensional gases. (b) Bragg spectrum of an array of strongly correlated 1D gases produced by lattices with amplitude $s = 50$. Red curve is a fitting of the function $\nu f(\nu)$, where $f(\nu)$ is a Lorentzian. (c) Half width at half maximum of the Bragg resonances as a function of the amplitude $s$ of the lattices, for two different value of momentum [blue squares ($q_1 = 16.0(2) \mu$m$^{-1}$) and black dots ($q_2 = 7.3(2) \mu$m$^{-1}$)].}
\label{Fig1}
\end{figure}

Our system, sketched in Fig.~\ref{Fig1}(a), consists of about 2$\times$10$^3$ 1D atomic micro-tubes. Each gas has typical total size $\sim$30 $\mu$m $\times$ 0.05 $\mu$m and linear density $n_{1D}\sim$5$\mu$m$^{-1}$. To arrange atoms in this configuration, we confine a 3D BEC of $^{87}$Rb in a pair of orthogonal red-detuned optical lattices. We study different configurations by tuning the amplitude $V$ of the 2D optical lattice ($s=V/E_R$ ranging from 0 to 56, where $E_R=h^2/2 m \lambda_L^2$, $h$ being the Planck's constant, $m$ the atomic mass and $\lambda_L = 830$nm the lattice wave-length).
The stronger the optical confinement, the more anisotropic is the trap experienced by each 1D gas (the aspect ratio $\lambda=\omega_\perp/\omega_\parallel$, namely the ratio of the radial harmonic trapping frequency to the axial one, ranges from 787 to 880 for $5 < s < 56$).
For all the amplitudes of the transverse lattice we explore, each gas has a fully 1D character, \emph{i.e.} both chemical potential and temperature are about one order of magnitude smaller than the frequency of the transverse harmonic oscillator. The crucial quantity to describe the regime of the 1D gas is the parameter $\gamma = m g_{1D}/\hbar^2 n_{1D}$, that is the ratio of interaction energy to the kinetic energy necessary to correlate particles at distance $1/n_{1D}$, $g_{1D}$ being the interatomic coupling in the 1D gas. In our case $\gamma \sim 0.3 \div 0.6$ so that inter-particle correlations are stronger than in the mean-field regime (see for example \cite{laburthetolra2004}).

We first investigate the effect of the phase-fluctuations via Bragg spectroscopy. In brief, the lattice gas is diffracted from a moving lattice created by two simultaneous off-resonant light pulses (Bragg beams) with a relative angle $\theta$, detuned from each other by a tunable frequency-difference $\omega / (2 \pi)$. This perturbs the system activating excitations with energy $\hbar \omega$ and momentum $\hbar \textbf{q}_B$, the modulus of which depends on $\theta$ \cite{stenger1999}. The geometry of the Bragg beams is chosen to align $\hbar \textbf{q}_B$ to the axis of the 1D tubes. After the excitation, the lattice amplitude is turned down in 15ms to a low value ($s = $ 5) where the different tubes are no more independent allowing the system to re-thermalize via atom-atom collisions. After 5ms both optical and magnetic trap are simultaneously switched off and the system is observed after a time of flight (TOF) $t_{ TOF} = 21$ms. The physical observable is the increase of the size of the central peak of the atomic cloud. More detailed comments on the experimental procedure have been reported in Ref.\cite{clement2009NJP}. Our technique allows to measure the  energy $\Delta E$ transferred to the system \cite{fabbri2009clement2009}, which depends on the imaginary part of its polarizability $\chi^{\prime \prime }(\omega)$, apart from the characteristics of the perturbing potential (amplitude $V_B$ and time duration $\Delta t$)  \cite{brunello2001}:  $\Delta E \propto V_B^2 \omega \chi^{\prime \prime }(\omega) \Delta t$. The polarizability can be expressed in terms of the dynamical structure factor of the system $\chi _{F}^{\prime \prime }(\omega)= \pi S(\omega,q)(1-e^{- \hbar \omega / (k_B T)})$.

Two different geometrical configurations of the Bragg beams have been used to vary the transferred momentum along the axis of the gases. Counter-propagating beams along the axis of the atomic tubes yield $q_1=16.0(2) \mu$m$^{-1}$; a small-angle configuration gives $q_2=7.3(2) \mu$m$^{-1}$. In both the cases we assume the excitation to be in the Doppler regime \cite{nota0}, where $S (q, \omega)$ is reduced to the momentum distribution $n (q)$ \cite{stenger1999,richard2003}. In this regime the spectral half-width at half-maximum (HWHM) can be related to the momentum width $\hbar \Delta q$ through the relation $\Delta \nu = \frac{ q_B}{2 \pi m} \hbar \Delta q$, which is linear in the wave-vector $q_B$ of the excitation \cite{nota1}. In the experiment, the ratio of the HWHMs of the response of identical arrays of 1D gases to the two different excitations $q_1$ and $q_2$ is consistent with $q_1/q_2 = (2.16 \pm 0.06)$, as expected (linear fitting of the experimental data on Fig.\ref{Fig1}(c) allows for defining a mean ratio $\Delta \nu_1/\Delta \nu_2=(2.7 \pm 0.8 )$).

In our range of $\gamma \approx 0.3 \div 0.6$, interactions are beyond the mean-field description but still far from the Tonks-Girardeau regime. Thus, we expect the interaction-induced spatial decay of one-particle correlation function to happen on a larger scale than that leaded by phase-fluctuations, for typical temperatures in the experiment ($\sim$100nK) \cite{priv}. The one-particle correlation function being dominated by the exponential decay due to phase-fluctuations, the momentum distribution exhibits a profile well described by a Lorentzian shape \cite{cazalilla2004petrov2000}. Its HWHM  $\hbar \Delta q$ relies only on the coherence length of the gas, being
\begin{equation}
\Delta q = \frac{0.635}{L_\phi} \label{Dp}
\end{equation}
where $L_\phi=\hbar^2 n_{1D} / (m k_B T)$ is the half coherence-length ($T$ being the temperature); the factor 0.635 take into account the 1D geometry \cite{gerbier2004}.

\begin{figure}[h!]
\includegraphics[width=0.8\columnwidth]{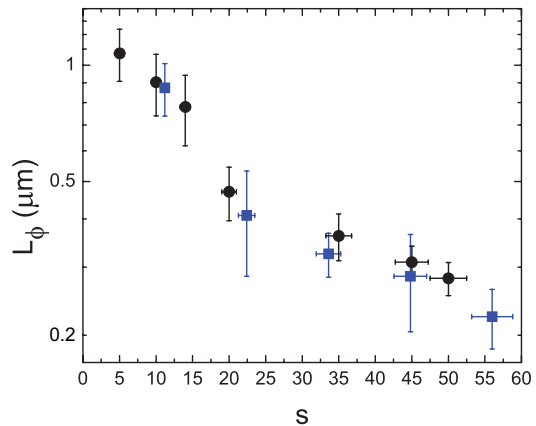}
 \caption{$L_{\phi}$ of the array of 1D gases is reported as a function of the amplitude $s$ of the lattices which squeezes the gas in 1D micro-tubes and for two different values of the momentum of the excitation [blue squared ($q_1$) and black dots ($q_2$)].}
\label{Fig2}
\end{figure}

The description of the problem is complicated by the presence of an array of gases with different densities (thus, different characteristic $L_\phi$). In principle one should consider the sum of the response of each tube. Sup\-po\-si\-ng a mean-field picture \cite{nota2}, interactions would give a broadening of the width and a shift of the center of the Bragg resonance compared to the single-particle response $h \nu_{s.p.} = \hbar^2 q^2/(2m)$, both depending on the density of each tube \cite{stenger1999}. Yet the global response of the system to the Bragg excitation consists in a single broad resonance, as depicted in Fig.~\ref{Fig1}(b), the center of which is shifted compared to $\nu_{s.p.}$ ($\nu_0>\nu_{s.p.}$). Its shape is well described by $\nu f(\nu) \simeq \omega S(q,\omega)$ where $f(\nu)$ is a Lorentzian function (see Fig.~\ref{Fig1}(b)). This suggests that thermal broadening of the response of each gas exceeds the interaction-induced broadening and masks the relative shifts of the resonant frequencies of the tubes. We verified numerically that it works for our experimental parameters \cite{articololungo}. Therefore we analyze the Bragg spectra as being the response of a single 1D gas and we define accordingly an \emph{effective} coherence-length $L_\phi$ of the whole system, using the relation in Eq.~\ref{Dp}.

From the fittings of the Bragg spectra, we extract the HWHM $\Delta \nu$. This quantity is reported in Fig.\ref{Fig1}(c) as a function
of the amplitude $s$ of the optical lattices and for the two different wave-vectors of the excitation $q_1,q_2$. The total number of atoms is kept almost constant in both the series of data. From the spectral half-width, we extract the half-length $L_{\phi}$. As shown in Fig.\ref{Fig2}, $L_{\phi}$ rapidly drops down as $s$ increases. In addition, the analysis of the spectra for the two momenta $q_1$ and $q_2$ reveals consistent $L_\phi$ as expected in the Doppler regime. The optical confinement makes the aspect ratio of the 1D gases grow and their density decrease. However, we estimate numerically \cite{articololungo} that the relative variation of the 1D-density in the whole range of $s$ is about 10\% and does not justify the rapid downfall of the coherence length. This suggests that major role in determining $L_\phi$ is played by the finite temperature. Note that extracting temperature from the measurement of $L_\phi$ is not straightforward as it requires to take into account the inhomogeneity of $n_{1D}$ over the array; this will be the subject of a future work \cite{articololungo}.

\begin{figure}[h!]
\includegraphics[width=0.7\columnwidth]{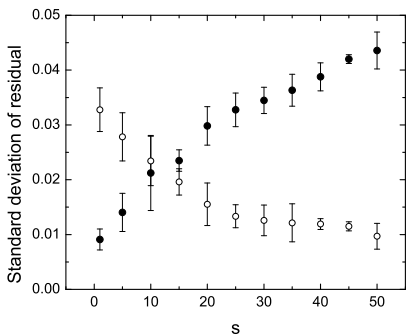}
 \caption{Mean-square value of residuals of a Lorentzian (hallow dots) and Gaussian function (solid dots) fitted to the momentum distribution mapped through TOF.}
\label{Fig3}
\end{figure}

To keep the insight up on the system, information on the coherence-length induced by thermal phase-fluctuations has been also extracted by direct mapping the momentum-distribution into space-distribution, which is measured via absorption imaging of the gas after switching-off the trap (below referred to as TOF measurements). The expansion of the atomic gas from the trap is governed by two kinds of kinetic-energy: the one which interactions convert into, and the one produced by in-trap phase-fluctuations. Due to the strong anisotropy of the trap, the interaction-induced expansion mainly affects the radial direction \cite{castindum1996}, whereas the longitudinal size of the cloud $R^{int}_{ TOF}$ is not significantly altered compared to its in-trap value. At non-zero temperature, thermally-induced local phase-gradients produce a velocity field given by $\textbf{v}_\phi = (\hbar / m) \nabla \phi $ \cite{pitaevskii}, $\phi$ varying significantly on a length-scale $L_\phi$. It determines an increase $R^\phi_{ TOF}$ of the longitudinal size during TOF, which contributes relevantly if $R^\phi_{ TOF}/R^{int}_{ TOF} > 1$, where
\begin{equation}
\frac{R^\phi_{ TOF}}{R^{int}_{ TOF}}\sim \frac{\hbar t_{TOF}}{m L_\phi R^{int}_{ TOF}}.
\label{ratio}
\end{equation}

In previous experiments, the product $L_\phi R^{int}_{ TOF}$ amounts typically to 10$\mu$m$\times$260$\mu$m for elongated 3D quasi-condensates \cite{richard2003} and $\sim$1$\mu$m$\times$170$\mu$m in the case of atom-chip experiments \cite{hofferberth20072008}. In both cases, $R^\phi_{ TOF}$ is negligible and the longitudinal length after TOF cannot be related to in-trap phase-fluctuations. For our 1D lattice gases, this quantity is reduced to $\sim 1 \mu$m$\times$27$\mu$m for $s = 5$ and it falls even one order of magnitude as the amplitude of the optical confinement increases ($\sim 0.2 \mu$m$\times$22$\mu$m for $s = 56$). Thus one expects that the in-trap size still dominates for low values of $s$ and the density profile has a parabolic shape (possibly smoothed to a Gaussian by the finite resolution of the imaging system), as expected in the Thomas-Fermi regime. For high values of $s$ phase-fluctuations should enlarge the distribution and the profile should assume a Lorentzian shape. To confirm this behaviour we have analyzed the TOF profiles with both Gaussian and Lorentzian functions. From these fittings we plot in Fig.~\ref{Fig3} the mean squared value of the residuals of both fitting functions. As anticipated from the simple formula in Eq.~\ref{ratio}, Fig.~\ref{Fig3} points out the cross-over through the two regimes at $s\sim10\div15$. In particular, for $s > 20$ initial trapped phase fluctuations are responsible for the Lorentzian shape of the TOF momentum distribution.

\begin{figure}[h!]
\includegraphics[width=0.69\columnwidth]{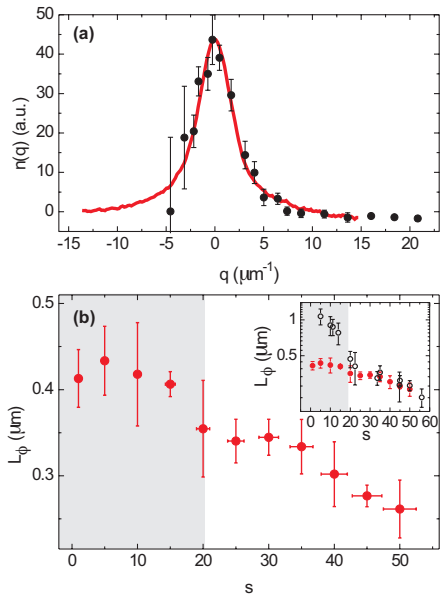}
\caption{(a) Momentum distribution of an array of 1D gases in an strongly confining optical lattice ($s = 50$) measured through Bragg spectroscopy (black dots) and direct mapping in TOF density profile (red curve). (b) Half coherence length $L_\phi$ extracted from TOF measurements is shown as a function of $s$ in linear scale. Gray area points out the region of parameters where TOF measurements can not be used to extract $L_\phi$. Inset: Comparison between $L_\phi$ from Bragg measurements (black dots) and direct mapping (red dots).}
\label{Fig4}
\end{figure}

To quantitatively compare the results of Bragg spectroscopy and TOF measurements, we map both energy spectra and density profile after TOF into momentum-space of the in-trap gas. In the first case, we use the free-particle dispersion relation obtaining $q = 4 \pi^2 m \nu /(h q_B) - q_B/2$. In the latter case, the calibration of the pixel size in momentum-space is obtained by measuring the distance between two interference peaks released from the lattices at weak amplitude. For an array of strongly-correlated 1D gases ($s = 50$ in Fig.\ref{Fig4}(a)) the momentum distributions measured via Bragg spectroscopy (data points) and TOF measurements (continuous line) show an excellent agreement. From TOF measurements we extract the coherence length as well. Accordingly to our results from Bragg spectroscopy, $L_\phi$ is observed to reduce as $s$ increases (Fig.~\ref{Fig4}(b)). In the inset, we compare $L_\phi$ measured in the two ways. The accordance fails in the region of parameters (gray area) where TOF profiles are not dominated by phase-fluctuations (see Eq.~\ref{ratio}).

In conclusion, we investigate the coherence properties of an array of 1D Bose gases by measuring their momentum distribution. We observe the latter to have a Lorentzian shape as predicted for a single uniform 1D gas. We define an effective coherence length of the whole ensemble and we show its evident reduction as the optical confinement is increased. Comparing Bragg spectroscopy and direct mapping of momentum into density distribution after TOF demonstrates that TOF images give access to coherence properties in the presence of strong phase-fluctuations. Our work paves the way for future studies of the coherence properties in 1D geometries with short coherence lengths, and especially strongly interacting disordered systems. Of particular interest are strongly interacting disordered systems where the   role of thermal phase fluctuations in the nature of the superconductor-insulator transition is debated \cite{phillips2003dubi2007}. So far, only disordered quasi-condensates have been investigated where it was shown that the contribution of phase fluctuations is small \cite{clement2008}.

We thank T. Giamarchi, A. Iucci and M. Zvonarev for fruitful discussions. This work has been supported by MIUR PRIN 2007, ECR Firenze, ERC-DISQUA Project, DQS EuroQUAM project from ESF, NAMEQUAM and AQUTE projects from EU. D.C. acknowledges support of a Marie Curie Intra-European Fellowship.


\begin{thebibliography}{99}
\bibitem{nota0} Doppler regime is described by the condition $q > \sqrt{4 \mu m} / \hbar$, where $\mu$ is the chemical potential of the gas.
\bibitem{nota1} To test the accuracy of this approximation, we calculate the HWHM of the momentum distribution for a single 1D gas from the spectral width using free-particle-like versus Bogoliubov dispersion relation and we verify that the results differ less than 10\%.
\end{thebibliography}
\end{document}